\documentstyle[aps,epsfig,amstex]{revtex}
\newcommand{\mel}[3]{\langle #1 | #2 | #3 \rangle } 
\newcommand{\ket}[1]{| #1 \rangle } 
\newcommand{\bra}[1]{\langle #1 | } 
\newcommand{\dt}{\Delta t}

\newcommand{\bfrho}{\mbox{\boldmath $\rho$}} 
\begin{document}
\title{Iterative algorithm versus analytic 
solutions of the parametrically driven dissipative quantum harmonic oscillator}
\author{Michael Thorwart, Peter Reimann, and Peter H\"anggi}
\address{
              Institut f\"ur Physik -
              Universit\"at  Augsburg, Universit\"atsstr.\ 1, 86135 Augsburg,
              Germany\\
}
\date{Date: \today}
\maketitle
\begin{abstract}
We consider the Brownian motion of a quantum mechanical particle 
in a one-dimensional parabolic potential with periodically modulated 
curvature under the influence of a thermal heat bath.
Analytic expressions for the time-dependent 
position and momentum variances 
are compared with results of an iterative algorithm, 
the so-called quasiadiabatic propagator 
path integral algorithm (QUAPI).
We obtain good agreement over an extended range of parameters for 
this {\em spatially continuous} quantum system.  
These findings indicate the reliability of the 
algorithm also in cases for which
analytic results may not be available {\em a priori}. \\[10mm]
\end{abstract}
\vspace{-2mm}
\section{Introduction}
Exactly solvable systems have a special status among physical models.
Although oversimplified in many cases, they may serve as starting point 
for testing the reliability of
methods which can then be transferred to more realistic, but only 
numerically
solvable models. An important class of such models are  
quantum systems coupled to a dissipative environment and being driven by
a time-dependent external field \cite{Weiss93}. A wide variety of physical 
phenomena
have been described by this kind of models, e.g.\ electron \cite{Marcus} 
and proton
\cite{Nagaoka} transfer, tunneling  processes of a macroscopic spin 
\cite{Wernsdorfer}, hydrogen
tunneling in condensed phases \cite{Wipf}, single defect tunneling in 
mesoscopic quantum 
wires \cite{Golding}, or tunneling of the magnetic flux in a SQUID 
\cite{Han}, to name but a few. Usually such dissipative quantum 
systems consist of 
 a model Hamiltonian bilinearly coupled to a bath
of harmonic oscillators. Additional external time dependent driving fields 
render the mathematical solution even more difficult or even impossible. 

One of the few analytically tractable time-dependent dissipative quantum 
systems is
the parametrically driven harmonic oscillator whose 
analytic solution was found by Zerbe and H\"anggi in Ref.\  \cite{Zerbe95}. 
A physical realization of this model is the Paul trap \cite{Brown91}, 
which provides an oscillating quadrupole potential for the enclosed ion. 
Furthermore, the parametrically driven dissipative harmonic oscillator may 
 serve as a benchmark for approximation schemes which were 
developed for more general dissipative systems \cite{Dittrich}. 
The interesting feature is that 
the parametric driving  induces a non-trivial quasienergy 
spectrum \cite{Dittrich}, 
in contrast to additive driving where the quasienergy spectrum 
coincides with the spectrum of the undriven system apart from a constant 
shift. This is further corroborated by the fact that the solution of the 
parametrically driven linear oscillator can be utilized to obtain 
solutions of certain nonlinear dynamical systems \cite{Lutzky78}. 

Powerful approximative numerical procedures for 
simulating dissipative and possibly time-dependent quantum systems 
 are the Quantum Monte Carlo method
\cite{Binder88} and 
the quasiadiabatic propagator path integral algorithm (QUAPI) developed 
by Makri and Makarov \cite{QUAPI}. The former method works very well for 
problems involving  
path integrals in imaginary time, however, the calculation of real 
time path integrals is afflicted
by the so called {\it sign-problem} due to the rapidly oscillating
integrand.  
The QUAPI algorithm has been applied to low dimensional dissipative
systems such as the  driven spin-1/2-particle coupled to a harmonic 
oscillator
bath (driven spin-Boson-system) \cite{QUAPI} and to the  driven 
double-well potential in order to study quantum hysteresis and quantum
stochastic resonance \cite{Thorw,qsr}. Moreover, the QUAPI algorithm has 
recently been used as a basis for a very efficient memory equation 
algorithm for spin-boson-models \cite{Golosov}. 

The purpose of this paper is to apply the QUAPI algorithm to the 
parametrically driven harmonic oscillator and to compare the 
results with the  analytic  solution from Ref.\  \cite{Zerbe95}. 
Whilst harmonic oscillator systems are known to exhibit some 
untypical features this is 
not the case with respect to the QUAPI algorithm. 
Our results thus show that not only intrinsically discrete models 
like the spin-boson-system but also {\it spatially continuous} 
systems can be
accurately described
by few energy eigenstates if the temperature is restricted to a 
moderate  regime. 
Most importantly,
 this is the first work in which the numerical approximative 
QUAPI results are
compared against {\it analytic} solutions of a  
{\it spatially continuous}  driven dissipative quantum system. 

The paper is organized as follows: In section \ref{spddho}, we 
introduce our model of the parametrically driven dissipative quantum 
harmonic oscillator and briefly review the analytic solution 
given in Ref.\ 
\cite{Zerbe95}. Section \ref{quapirev} is
devoted to a short review of the QUAPI method.  The comparative 
main results are
presented in section \ref{results}, before we give the conclusions  
in section \ref{conclusio}. \\
\section{The model and its analytic solution}
\label{spddho}
In this section we briefly review  the analytic solution of the 
parametrically driven dissipative harmonic oscillator from \cite{Zerbe95}. 

A quantum particle with mass $M$, position operator ${\bf x}$ and 
momentum operator ${\bf p}$  moving in a one-dimensional harmonic 
potential with  periodically modulated curvature is described by 
the Hamiltonian
\begin{equation}
{\bf H}_{\rm S} (t)= \frac{{\bf p}^2}{2M} + \frac{M}{2} [ \omega_0^2 + 
\epsilon \cos \Omega t ] {\bf x}^2. \label{system}
\end{equation}
Following the  common approach \cite{Weiss93} to model the influence 
of the environment by an ensemble of harmonic oscillators, the bath 
Hamiltonian ${\bf H}_{\rm B}$ (including the interaction with the 
system) is given by 
\begin{equation}
{\bf H}_{\rm B}=\sum_j {\bf H}_j ({\bf x}) = \sum_j \frac{1}{2}\Big 
[\frac{{\bf p}_j^2}{m_j}+m_j\omega_j^2
\Big ({\bf q}_j-\frac{c_j}{m_j\omega_j^2} {\bf x} \Big ) ^2 \Big ], 
 \label{hamilton}
\end{equation}
and the whole system is described by the Hamiltonian 
${\bf H}(t)={\bf H}_{\rm S}(t)+{\bf H}_{\rm B}$. In the case of a 
thermal equilibrium bath, it turns out that its influence on the 
system is fully characterized by the spectral density 
\begin{equation}
J(\omega)=\frac{\pi}{2}\sum_j \frac{c_j^2}{m_j \omega_j} \delta(\omega - \omega_j).
\end{equation}
With the number of harmonic oscillators going to infinity, we 
arrive at a continuous spectral density. In the following, we choose 
for the sake of definiteness a truncated  Ohmic spectral density, i.e.\  
\begin{equation}
J(\omega)=M \gamma \, \omega f_c(\omega,\omega_c). \label{specdens}
\end{equation}
Here, $\gamma$ is the coupling strength to the heat bath and 
$f_c(\omega,\omega_c)$ denotes a cut-off function which avoids 
unphysical divergences due to high-frequency bath modes. For our 
calculations, we consider two examples for the cut-off function: 
(i) a smooth exponential cut-off 
\begin{equation}
f_c(\omega,\omega_c)=\exp (-\omega/\omega_c) \label{cut1}
\end{equation}
and (ii) a step-function 
\begin{equation}
f_c(\omega,\omega_c)=\Theta (\omega_c - \omega) \label{cut2}
\end{equation}
with cut-off frequency $\omega_c \gg \omega_0,\Omega$ 
(see discussion given below). 

We choose a factorizing initial condition of Feynman-Vernon form  
\cite{Feynman63} which means that at 
time $t=t_0$, the full density operator ${\bf W}(t_0)$ is given as 
a product of the initial 
system density operator ${\bfrho}_{\rm S}(t_0)$
and the canonical bath density operator
at temperature $T=1/k_B \beta$, i.e.,
\begin{equation}
{\bf W}(t_0) = {\bfrho}_{\rm S}(t_0)\, \, Z_{\rm B}^{-1}\, 
\exp (-\beta {\bf H}_{\rm B}^0)\ ,
\label{ic}
\end{equation}
where $Z_{\rm B}^{-1} = Tr \exp (-\beta {\bf H}_{\rm B}^0)$ and  
\begin{equation}
{\bf H}_{\rm B}^0 =
\sum_j \frac{1}{2}\Big [\frac{{\bf p}_j^2}{m_j}+m_j\omega_j^2\, {\bf q}_j ^2 \Big ]\ . 
\label{hamilton0}
\end{equation}
By way of integrating out the bath degrees of freedom in Eq.\  
(\ref{hamilton}) one 
obtains the following one-dimensional Heisenberg equation for the 
position operator ${\bf x}$, i.e.\  
\begin{equation}
\ddot{{\bf x}}(t) + \int_{t_{0}}^t 
\hat{\gamma} (t-t^\prime )\dot{\bf x}(t^{\prime})dt^{\prime} + 
(\omega_0^2 + \epsilon \cos \Omega t){\bf x}(t) = \frac{1}{M} {\bf \Gamma}(t)
 - \hat{\gamma} (t-t_0 )\, {\bf x}(t_0)\ , 
\label{langevin}
\end{equation}
with the friction kernel given by 
\begin{equation}
 \hat{\gamma} (t) = \frac{2}{M \pi} \int_0^\infty d\omega 
 \frac{J(\omega)}{\omega} \cos(\omega t). 
\label{dampkern}
\end{equation}
${\bf \Gamma}(t)$ is a time-dependent fluctuating (operator) force 
\begin{equation} 
{\bf \Gamma}(t)= 
\sum_j c_j\, \left( 
\frac{{\bf p}_j(t_{0})}{m_j \omega_j} \sin (\omega_j (t-t_0)) 
+ {\bf q}_j(t_{0}) \cos (\omega_j (t-t_0)) \right) 
\label{3.3a} 
\end{equation} 
which contains the initial conditions of the bath and of the 
particle's position at time $t_{0}$. 
The last term on the r.h.s.\  (proportional to ${\bf x}(t_{0})$) 
in Eq.\   (\ref{langevin}) is the so-called initial slip, 
caused by the specific choice (\ref{ic}) of the initial conditions.

Exploiting the thermal 
distribution of the bath one recovers the usual connection 
(via $J(\omega )$) 
between the random and the frictional forces of the bath 
in Eq.\  (\ref{3.3a}) 
in the form of the fluctuation-dissipation-relation, reading, 
$t \ge t'$,  
\begin{equation} 
\langle {\bf \Gamma}(t) {\bf \Gamma} (t') \rangle_{\beta} 
= Tr \left [ Z_{\rm B}^{-1} \exp (-\beta {\bf H}_{\rm B}^0) \, 
{\bf \Gamma}(t) {\bf \Gamma} (t') \right ] 
= \hbar L(t-t')\ , 
\label{L} \ 
\end{equation} 
\begin{equation} 
L(t)={1 \over \pi}\int_0^\infty d\omega J(\omega ) \left [ 
\mbox{coth} \left ( {\hbar \omega \beta \over 2} \right ) \cos (\omega 
t) - i\sin( \omega t)\right ] , 
\label{2.4} 
\end{equation} 
where  the subscript $\beta$ indicates thermal 
averaging performed with the canonical density operator 
for ${\bf H}_{\rm B}^0$ defined in Eq.\  (\ref{hamilton0}). 
The response function $L(t)$ 
will play an important role in the numerical QUAPI algorithm. 
%
%

It turns out \cite{Zerbe95} that for the description of the parametric 
dissipative quantum 
oscillator the solution of the 
classical {\it deterministic} limit ($\hbar \rightarrow 0, T \rightarrow 0$) 
with $\omega_c \rightarrow 
\infty$ plays a prominent role. Thus, in Eq.\ (\ref{langevin}) the 
position operator $\bf{x}$ is 
replaced by the classical coordinate $x$ and $\int_{t_{0}}^t 
\hat{\gamma} (t-t^\prime )\dot{\bf x}(t^{\prime})dt^{\prime}$ goes 
over into 
$\gamma \dot{x}(t)$. Moreover, on the right hand side of Eq.\ (\ref{langevin}),
the fluctuations ${\bf \Gamma }(t)$ are zero and the initial slip 
is also omitted, which
can be achieved by either a somewhat different choice of the 
initial conditions than
in Eq.\  (\ref{ic}) or by replacing the coupling coefficients $c_j$ in Eq.\  
(\ref{hamilton})
by $c_j\, \Theta (t-t_0^+ )$ so that ${\bf H}_{\rm B}$ and ${\bf H}_{\rm B}^0$ 
from Eq.\  (\ref{hamilton0})
coincide at $t=t_0$.
For convenience we furthermore introduce scaled quantities 
%
\[
\begin{array}[t]{cclccclcccl}
\tilde{t}&=& \frac{\Omega}{2} t, & &
\tilde{x}(\tilde{t})&=& \sqrt{M \Omega / 2 \hbar} \, x 
(t=\frac{2 \tilde{t}}{\Omega}), & & 
\tilde{\omega}_{0} &=&
 \frac{2}{\Omega} \omega_{0}, 
\end{array} 
\]
\begin{equation} 
\begin{array}[t]{cclccclccclcccl}
\tilde{\epsilon} & = &\frac{2}{\Omega^2} \epsilon, & & \tilde{\gamma} 
&=& \frac{2}{\Omega}\gamma , & &\tilde{T} &=& \frac{2 k_B }{\hbar \Omega} T, 
& & \tilde{\omega}_{c} &=&
 \frac{2}{\Omega} \omega_{c}.
\label{scale}
\end{array} 
\end{equation}
In the remainder of this paper, {\it we exclusively use 
dimensionless quantities} but omit all the tildes for the sake 
of better readability. In order to recover the dimensionful quantities, 
one has to re-introduce tildes  
wherever it makes sense and then exploit Eq.\  (\ref{scale}).
By substituting  
$x(t)=y(t) \exp[-\gamma ( t-t_0) /2]$ we  arrive at 
an undamped oscillator equation for $y$  which  is the 
well-known Mathieu equation 
\begin{equation}
\ddot{y} (t) + (\omega_0^2 - \frac{\gamma^2}{4} + 2 \epsilon \cos 2 t) 
y(t) = 0\ . \label{mathieu}
\end{equation}
Its mathematical properties like stability and instability 
regions in the parameter space are well known \cite{mathieueq}. 
Nevertheless, there exists no closed analytic expression for 
the solution and the equation has to be integrated numerically.
 In the following, we will need two linear independent 
 solutions $\Phi_i(t), i=1,2$, of Eq.\  (\ref{mathieu}) belonging to 
 two different sets of initial conditions
\begin{equation}
\begin{array}[t]{cclcccl}
\Phi_1(t_0)&=& 0, & & \dot{\Phi}_1 (t_0)&=& 1, \nonumber \\
\Phi_2(t_0)&=& 1, & & \dot{\Phi}_2 (t_0)&=& 0 \, \label{iniphi}.
\end{array} 
\end{equation}
They can be determined numerically, e.g.\ by means of a regular 
fourth-order Runge-Kutta integration of the Mathieu equation 
(\ref{mathieu}). 

Let us return to the dissipative quantum parametric oscillator. 
The quantities of interest are the variances of the position 
and the momentum operator, i.e.\
\begin{eqnarray}
\sigma_{xx}(t) & \equiv & \langle {\bf x}^2(t) \rangle - \langle {\bf x}(t) 
\rangle^2, \nonumber \\
 \sigma_{xp}(t) & \equiv & \frac{1}{2}\langle {\bf x}(t) {\bf p} (t) + 
 {\bf p}(t) {\bf x} (t)  
 \rangle - \langle {\bf x} (t) \rangle \langle {\bf p}(t) \rangle, \nonumber \\
\sigma_{pp}(t) & \equiv & \langle {\bf p}^2(t) \rangle - \langle {\bf p}(t) 
\rangle^2.
\label{basic}
\end{eqnarray}
Here, the quantum mechanical expectation value is understood as 
usual as $\langle \cdot \rangle = Tr [\rho(t) \cdot ]$. 
By determining the propagator ${\bf U}(t,t_0) = {\cal T} 
 \exp(-i \int_{t_0}^t dt' {\bf H} (t') / \hbar)$ 
 (${\cal T}$ is the time ordering operator) for the driven dissipative 
system according to \cite{Zerbe95}, 
the reduced density matrix ${\bfrho}(t) = Tr_{Bath} ({\bf U}(t,t_0) 
{\bf W} (t_0) 
 {\bf U}^{-1}(t,t_0))$ can be calculated analytically. Here,
 ${\bf W}(t_0)$ denotes the full density 
 operator at time $t_0$ and $Tr_{Bath}$ the trace over the bath 
 degrees of freedom. Having obtained the reduced density operator 
 ${\bfrho}(t)$, the quantum mechanical expectation values in Eq.\  
 $\ref{basic}$) can 
 be evaluated. 
 After some algebra, we find for the dimensionless  
variances the expressions
\begin{mathletters}
\label{final}
\begin{eqnarray}
\sigma_{xx} (t) & = & e^{-\gamma (t-t_0)} \{[\Phi_2(t) - 
\frac{\gamma}{2} \Phi_1(t)]^2 \sigma_{xx}^0 
+ 2 \Phi_1(t) [\Phi_2(t) - \frac{\gamma}{2}\Phi_1(t)] 
\sigma_{xp}^0 + \Phi_1^2(t) \sigma_{pp}^0 \}+ \Sigma_{xx}(t), 
\label{final1}
\\ 
\sigma_{xp} (t) & = & \frac{1}{2} \dot{\sigma}_{xx} (t), 
\label{final2}
\\
\sigma_{pp} (t) & = & \dot{\sigma}_{xp}(t) + \gamma \sigma_{xp}(t) + 
[\omega_0^2 + 2 \epsilon \cos (2 t)] \sigma_{xx}(t) - \Sigma_{pp}(t). 
\label{final3}
\end{eqnarray}
\end{mathletters}
Thereby, we have rectified \cite{misprints} some minor misprints 
in \cite{Zerbe95} and simplified 
the equations in \cite{Zerbe95} for $\sigma_{xp}$ and $\sigma_{pp}$. Here, 
$\sigma_{xx}^0,\sigma_{xp}^0$ and $\sigma_{pp}^0$ denote the 
initial variances of the {\it uncoupled} system at time 
$t=t_0$ which depend on the choice of the initial state for the 
bare system ${\bf H}_{\rm S} (t_0)$. The initial conditions for 
Eqs.\ (\ref{final}) at time $t=t_0^+$ are given by 
\begin{eqnarray}
\sigma_{xx} (t_0^+) & = & \sigma_{xx}^0, \nonumber \\
\sigma_{xp} (t_0^+) & = & -\gamma \sigma_{xx}^0 + \sigma_{xp}^0, 
\nonumber \\
\sigma_{pp} (t_0^+) & = & \gamma^2 \sigma_{xx}^0 -2 \gamma  
\sigma_{xp}^0+\sigma_{pp}^0. 
\label{inistep}
\end{eqnarray}
The discontinuity of the variances at time $t_0$ is a well-known 
consequence \cite{Weiss93}
of the initial slip term in Eq.\  (\ref{langevin}); it is 
due to the factorizing 
initial condition (\ref{ic}).   
The first terms in the three equations (\ref{final}) 
 posses the same form  as in the classical case. 
The specific quantum mechanical features enter via the functions 
$\Sigma_{xx}(t)$ and $\Sigma_{pp}(t)$, which read  
\begin{mathletters}
\label{variances}
\begin{eqnarray}
\Sigma_{xx}(t) & = & \frac{\gamma}{\pi} \int_{0}^{\infty} 
d\omega \, \omega f_c(\omega,\omega_c)  
\coth(\frac{\omega}{2 T}) \Bigg\{ 
\Big[ \int_{t_0}^{t} ds \, G(t,s) \exp (\frac{\gamma}{2}(t-s)) 
\cos(\omega s)   \Big]^2 
 \nonumber
 \\  & & 
 \mbox{}+ \Big[\int_{t_0}^{t} ds \, G(t,s) \exp (\frac{\gamma}{2}(t-s)) 
 \sin(\omega s)\Big]^2  \Bigg\}, 
 \label{variances1}
 \\
\Sigma_{pp}(t) & = & \frac{\gamma}{\pi} \int_{0}^{\infty} d\omega \, 
\omega f_c(\omega,\omega_c)   
\coth(\frac{\omega}{2 T}) 
\int_{t_0}^{t} ds \, G(t,s) \exp(\frac{\gamma}{2}(t-s)) \cos(\omega (t-s)) \, ,
\label{variances2}
\end{eqnarray}
\end{mathletters}
where $G(t,s)=\Phi_1(t) \Phi_2(s) - \Phi_1(s)\Phi_2(t)$. While 
in Eq.\  (\ref{variances}) a general form of the 
cut-off function $f_c(\omega, \omega_c)$ with  
$\omega_c \gg \omega_0$   
is kept, the analytic solution (\ref{final}) is 
based \cite{Zerbe95} on the assumption of a strictly Ohmic 
classical dynamics ($\omega_c \rightarrow \infty$) in 
Eq.\  (\ref{dampkern}). 
The consequence of this assumption 
is the discontinuity at $t=t_0$ in Eq.\  (\ref{inistep}) when the 
system-bath-interaction is switched on instantaneously. 
A finite cut-off in the spectral density $J(\omega)$ in 
the damping kernel (\ref{dampkern}) would induce a smoothend  
time evolution of the variances (\ref{final}) close to $t=t_0$ 
on a time scale $\omega_c^{-1}$. 

The relations (\ref{final}, \ref{variances}) are evaluated 
by standard numerical methods. The efficiency is improved if one 
applies Floquet's theorem for the fundamental solutions $\Phi_j(s)$. Then, 
the periodic part of the Floquet solutions can be expanded in a Fourier 
series and 
the integrations over the intermediate times $s$ in Eq.\  (\ref{variances}) 
can be performed analytically. 
Finally, the 
remaining  $\omega$-integrations and the sum over the Fourier modes 
can be readily carried out. 
\section{Numerical solution with real-time path integrals} \label{quapirev}
In the following section, we recapitulate the essentials of the 
QUAPI algorithm. Further  
details can be found in the original works by Makri and 
Makarov \cite{QUAPI}. 
In order to describe the dynamics of the system of interest 
it is sufficient to consider the time evolution of the elements 
of the reduced 
density matrix which reads in position representation 
\begin{eqnarray} 
\rho (x_f,x_f';t_f)&=&Tr_{Bath} 
\langle x_f \Pi_j q_j | {\bf U}(t_f,t_{0})\, {\bf W}(t_{0})\, {\bf U}^{-1} 
(t_f,t_{0}) | x_f' \Pi_j q_j' \rangle \ , \nonumber \\ 
{\bf U}(t_f,t_{0})&=&{\cal T}\exp \left \{ -i/\hbar \int_{t_{0}}^{t_f} {\bf H}(t') 
 dt' \right \} \  . 
\label{3.1new} 
\end{eqnarray} 
Here, ${\cal T}$ denotes the chronological operator, ${\bf W}(t_{0})$ the 
full density operator at the initial time $t_{0}$ and $Tr_{Bath}$ the 
partial trace over the harmonic bath oscillators $q_j$. Due to our 
assumption 
that the bath is initially at thermal equilibrium and decoupled from 
the system, ${\bf W}(t_{0})$ becomes the product of the 
initial system density operator ${\bfrho}_{\rm S}(t_{0})$ 
and the canonical bath density operator 
at temperature $T$, see Eq.\  (\ref{ic}). Then, the partial trace over the 
bath can be performed 
and the reduced density operator be rewritten according to 
Feynman and Vernon\cite{Feynman63} as 
\begin{equation} 
\rho (x_f,x_f^\prime ,t_{f}) = \int dx_0 dx_0^\prime \ G((x_f,x_f^\prime,t_{f} 
;x_0,x_0^\prime,t_{0}) \rho (x_0, x_0^\prime, t_{0}) \ ,
\label{2.1} 
\end{equation} 
with the propagator $G$ given by 
\begin{equation} 
G(x_f,x_f^\prime, t_{f} ;x_0,x_0^\prime,t_{0}) = \int {\cal D}x {\cal D}x^\prime 
\exp \left \{ {i\over \hbar} \left ( S_{\rm S}[x] - S_{\rm S}[x^\prime ] \right 
)\right \} {\cal F}_{FV}[x,x^\prime ]  . 
\label{2.2} 
\end{equation} 
$S_{\rm S}[x]$ is the classical action functional of the 
system-variable $x$ along a path $x(t)$ and 
${\cal F}_{FV}[x,x^\prime ]$ denotes the Feynman-Vernon influence functional 
\begin{equation} 
{\cal F}_{FV}[x,x^\prime ] = 
\exp \left\{  -{1 \over \hbar}\ 
\int_{t_{0}}^{t_{f}} dt 
\int_{t}^{t_{f}} dt' \left[ x(t') - x'(t') \right] 
\left [\eta(t' - t)x(t) - \eta^*(t' - t)x'(t) \right] 
 \right\} \ ,
\label{infl} 
\end{equation} 
with the integral kernel 
\begin{equation} 
\eta(t) = L(t)+ i \delta(t)\, \frac{2}{\pi} 
\int_0^\infty\frac{d\omega}{\omega}\, J(\omega) 
\label{2.4a} 
\end{equation} 
and the autocorrelation function $L(t)$ given in  Eq.\  (\ref{2.4}). 
As usual, the restriction to paths that satisfy 
the boundary conditions $x_0(t_{0})=x_0$, $x_f(t_{f})=x_f$ and similarly 
for $x'(t)$ 
is understood implicitly in Eq.\  (\ref{2.2}). Likewise, the dependence of the 
density operator ${\bfrho}$ in Eq.\  (\ref{2.1}) on the initial time $t_{0}$ and 
on ${\bfrho}_{\rm S}(t_{0})$ has been dropped. 

To make the equations numerically tractable, we discretize $t_{f}-t_{0}$ 
into 
$N$ steps $\Delta t$, such that $t_k = t_0 + k \Delta t$ and split  the 
full propagator over 
one time step ${\bf U}(t_{k+1},t_k)$ in Eq.\  (\ref{3.1new}) according 
to the Trotter formula symmetrically into a system and an environmental part: 
\begin{eqnarray} 
{\bf U}(t_{k+1},t_k) & \approx & \exp (-i {\bf H}_{\rm B} \Delta t 
/2 \hbar) {\bf U}_{\rm S}(t_{k+1},t_k) \exp (-i {\bf H}_{\rm B} \Delta t
/2 \hbar) \ , \nonumber\\ 
{\bf U}_{\rm S}(t_{k+1},t_k) & = & {\cal T} \exp \left \{ -\frac{i}{\hbar} 
\int_{t_k}^{t_{k+1}} dt'
{\bf H}_{\rm S}(t') \right \} \ .
\label{split1} 
\end{eqnarray} 
The symmetric splitting of the propagator in Eq.\  (\ref{split1}) 
causes an error proportional 
to $\Delta t ^3$. This error will be studied in detail in section 
\ref{results} below. The short-time propagator ${\bf U}_{\rm S}$ of 
the bare system 
is numerically evaluated by means of a Runge--Kutta 
scheme with adaptive step-size control. Exploiting the approximation
(\ref{split1}),  
the propagator in the position representation now factorizes 
 as 
\begin{equation} 
\langle x\, \Pi_j q_j |  {\bf U}(t_{k+1},t_k) | x'\,\Pi_j q'_{j} \rangle  \approx 
\langle x | {\bf U}_{\rm S}(t_{k+1},t_k) | x' \rangle \ 
\prod_{j} \langle q_j | e^{-i {\bf H}_j (x)  \Delta t/2\hbar} e^{-i {\bf H}_j (x') 
\Delta t/2\hbar} | q'_{j} \rangle \ , 
\label{quap} 
\end{equation} 
where the ${\bf H}_j(x)$ are defined in Eq.\  (\ref{hamilton}). 
By exploiting this approximation and performing the partial trace over 
the bath modes in Eq.\  (\ref{3.1new}), one recovers Eq.\  (\ref{2.1}), 
but now with a discretized version of the propagating function (\ref{2.2}), 
i.e.,   
\begin{eqnarray} 
\rho (x_f,x_f';t_{f}) & = & \int dx_0 \ldots \int dx_{N} 
\int dx'_0 \ldots \int dx'_{N}\,\delta(x'_f- x'_N)\,\delta(x_f- x_N)\, 
\nonumber \\ 
& & \times  \mel{x_N}{{\bf U}_{\rm S}(t_{f},t_{f}-\Delta t)}{x_{N-1}} 
\ldots \mel{x_1}{{\bf U}_{\rm S}(t_{0} +\Delta t, t_{0})}{x_0} 
\nonumber \\ 
& & \times  \mel{x_0}{{\bfrho}_{\rm S}(t_{0})}{x'_0} \mel{x'_0}{{\bf U}_{\rm S}^{-1} 
(t_{0}+\Delta t ,t_{0})}{x_1'} \ldots \mel{x'_{N-1}} 
{{\bf U}_{\rm S}^{-1}(t_{f},t_{f}-\Delta t)}{x'_N} \nonumber\\ 
& & \times  {\cal F}_{FV}^{(N)} 
(x_0,x'_0,\ldots ,x_{N},x'_{N}) \ . 
\label{quapi} 
\end{eqnarray} 
Here, ${\cal F}_{FV}^{(N)} (x_0,...,x'_N)$ 
is the discrete Feynman--Vernon influence functional (\ref{infl}) 
where the paths $x(t)$ and $x'(t)$ consist of constant segments 
$x_k$ and $x'_k$, respectively, within each time interval 
$t_k-\frac{1}{2} \Delta t < t_k < t_k+\frac{1}{2} \Delta t$ 
and can be rewritten in the form
\begin{equation} 
{\cal F}_{FV}^{(N)}(x_0,...,x'_N) = 
\exp\left\{-\frac{1}{\hbar}\sum_{k=0}^N\sum_{k'=k}^N 
[x_{k'}-x'_{k'}] [\eta_{k'k}x_{k}-\eta^*_{k'k}x'_{k}]\right\}\ . 
\label{discr1} 
\end{equation} 
The coefficients $\{\eta_{k'k}\}$ are closely related to 
their continuous time 
counterpart $\eta(t)$ in Eq.\  (\ref{2.4a}). Their explicit form is lengthy 
and not very illuminating for our purposes; their detailed form 
 can be looked up in Ref.\  \cite{QUAPI}. 
 
%
To make further progress, it is necessary to approximately break  
the influence kernel 
${\cal F}_{FV}^{(N)}(x_0,...,x'_N)$ in Eq.\  
(\ref{discr1}) into smaller pieces. To this end, Makri and Makarov  
use the fact  \cite{Leggett83,QUAPI} that 
the real part of the integral kernel $L_R(t)$ typically exhibits a 
pronounced peak at $t=0$, and 
quickly approaches $0$ for $t \rightarrow \pm \infty$. 
The decay to zero depends naturally on the choice of the cut-off function 
$f_c(\omega, \omega_c)$, see Eq.\  (\ref{specdens}). 
This suggests the truncation of $\eta(t)$ after a certain number $K$ of 
time steps $\dt$ and, correspondingly, to neglect $\eta_{k'k}$ if 
$k'>k+K$, i.e.
\begin{equation} 
{\cal F}_{FV}^{(N)}(x_0,...,x'_N) \approx 
\prod_{k=0}^N \prod_{k'=k}^{\min\{ N,k+K\}} 
\exp\left\{-\frac{1}{\hbar}[x_{k'}-x'_{k'}] 
[\eta_{k'k}x_{k}-\eta^*_{k'k}x'_{k}]\right\}\ . 
\label{discr2} 
\end{equation} 
In doing so, we approximate $L(t)$ by zero for $t > K \Delta t$,  
 cf.\  Eqs.\  (\ref{2.4}, \ref{2.4a}). Of course, this truncation induces an 
 error in the final result which has to be handled with care. 
 The error becomes increasingly less important for increasing temperatures 
 since then, the bath-induced correlations fall off increasingly faster. 
 In other words, for higher temperatures the width 
of the response function $L(t)$ decreases. In the other limit of 
decreasing temperature however, the number $K$ of relevant time-intervals 
is increasing and in the limit of 
zero temperature $T=0$, it is well known \cite{Weiss93} that the response 
function $L(t)$ falls off only algebraically for 
$t \rightarrow \pm \infty$. Nevertheless, we will see that this approach 
allows to deal with quite low temperatures and produces qualitative 
agreement with analytic solutions. 
 
The next goal is to approximate the spatially continuous integrals in Eq.\  
(\ref{quapi}) in terms of finite sums. To this end, 
 Makri and Makarov  perform a transformation into a basis 
given by the energy eigenstates $|\phi_m\rangle$ of 
the bare system Hamiltonian ${\bf H}_{\rm S}(t_r)$ (\ref{system}), but 
with the driving term clamped to an appropriate but fixed (!) 
reference time $t_r$, i.e.\ 
\begin{equation} 
{\bf H}_{\rm S}(t_r) \ket{\phi_m} = E_m \ket{\phi_m}\ ,\ m=1,2,... \  . 
\label{basistrans1} 
\end{equation} 
$E_m$ denotes the energy eigenvalues of the 
static system Hamiltonian ${\bf H}_{\rm S}(t_r)$. 
In certain cases, symmetry properties suggest the 
choice of an appropriate $t_r$. Here, we choose the 
unperturbed harmonic oscillator as a reference configuration. 
This means for our choice of driving to use  
$t_r = \pi/4$, so that $\cos (2t_r) = 0$ in Eq.\  (\ref{mathieu}). 
Reintroducing now the thermal bath but restricting ourselves to 
small-to-moderate 
temperatures $T$, the thermal occupation of high energy levels $E_m$ 
is expected to 
be negligible. This argument suggests that the $\ket{\phi_m}$ 
provide a 
well adapted basis admitting a fast convergent truncation scheme. 
In other words, 
we may approximately project the dynamics onto the Hilbert subspace 
spanned by the first few 
energy eigenstates $\ket{\phi_m}$, $m=1,...,M$, corresponding to an 
approximate 
decomposition of the identity operator ${\mathbb I} \approx 
\Sigma_{m=1}^M \ket{\phi_m}\bra{\phi_m}$. 
Before doing so, we perform one more unitary transformation within 
that $M$-dimensional Hilbert space such that the position operator 
becomes diagonal [discrete variable representation, DVR \cite{Harris}]: 
\begin{equation} 
\ket{u_m} = \sum_{m'=1}^M R_{mm'} \ket{\phi_{m'}}\ ,\ 
\mel{u_m}{{\bf x}}{u_{m'}} = x_m^{DVR} \delta_{mm'} \ , m,m'=1,..., M\ . 
\label{basistrans2} 
\end{equation} 
Exploiting the approximate decomposition of the identity 
${\mathbb I} \approx \Sigma_{m=1}^M \ket{u_m}\bra{u_m}$  and 
the truncation of the bath-induced correlations in Eq.\  (\ref{discr2}), 
it is a matter of 
straightforward but tedious manipulations -- starting from Eq.\  (\ref{quapi}) 
-- to arrive 
at the final form of the QUAPI recursion scheme. In particular, the 
integrals in Eq.\  (\ref{quapi}) turn into finite sums due to the transformation 
(\ref{basistrans2}) into the DVR-basis. We do not present the 
detailed form here and refer the reader again to 
the original literature \cite{QUAPI}. 

The above introduced restriction to a finite dimensional subspace induces 
an error in the evaluation of the reduced density matrix. However, as 
we will also discuss below, this error behaves in a controlled way if 
the relevant parameters such as the temperature and the damping 
 are chosen in a moderate 
regime. This means that for increasing temperature increasingly more 
DVR-states are necessary to describe the dynamics appropriately. 
Note that in this regime however, the number $K$ of the relevant 
memory time steps is decreasing. 
In the opposite limit of 
decreasing temperature, the number $M$ of relevant basis states can 
be chosen rather small. In this low-temperature limit the number $K$ 
of memory time steps can therefore be increased. Moreover, we note 
that the restriction of the dynamics (at long times) to the $M$-dimensional
 Hilbert subspace is not allowed for systems with an inherent diverging 
 dynamics. This is also seen in our example of the parametrically 
 driven dissipative quantum harmonic oscillator for a parameter 
 choice in an instability region of the Mathieu equation 
 (\ref{mathieu}), see section \ref{results} below. 

The efficiency of the QUAPI algorithm is based on the choice of the 
two free parameters $M$ (the number of basis states) and $K$ (the 
length of the memory). The numerical objects that one has to deal 
with are arrays of size $M^{2K+2}$ and $M^{2K}$. In practice, the 
calculations have been performed on conventional IBM RS/6000 workstations 
(43P-260 and 3CT). The computation time for an iteration over a typical time 
span $[0,40]$ depends strongly on the chosen parameters. It ranges from 
several milliseconds for $M=3, K=2$ (program size 7 MB) over several seconds 
for $M=3, K=4$ (program size 8 MB) up to several hours for $M=5, K=4$ 
(program size 176 MB). The strongly limiting factor is the program size 
since the size of the arrays grows exponentially with $K$. E.g., the 
parameter combination $M=4, K=6$ leads to too large arrays and cannot 
be treated by standard programming techniques. 
In practice, with the choice $M=6, K=3$ or $M=5, K=4$, we already are 
at the upper limit of the QUAPI algorithm.   
%
\section{Results}  \label{results}
We proceed in reporting our results for the specific 
example of the parametrically driven dissipative quantum harmonic 
oscillator. With the reduced density matrix (\ref{quapi}) at 
hand, we can calculate the variances (\ref{basic}) within 
the QUAPI
 algorithm and compare them with the analytic predictions 
 (\ref{final},\ref{variances}). 
 Most of the figures contain results for rather extreme parameter 
 values, e.g.\ low temperature and large driving amplitude, in 
 order to show that the QUAPI algorithm performs satisfactorily 
 also in these limits. For more moderate choices of the parameters, 
 the agreement (not shown) 
between numerical and analytic results is much better. 
 
Our main goal is to study the dependence of the variances 
(\ref{basic}) 
on the QUAPI parameters $M,K$ and $\Delta t$. For finite $M$ and $K$, 
the deviation increases proportional to $\Delta t^3$ due to the 
Trotter splitting in Eq.\  (\ref{split1}) with increasing $\Delta  t$. 
For decreasing $\Delta t$, the Trotter error decreases but the 
error made by the memory truncation in Eq.\  (\ref{discr2}) starts to 
dominate since more and more bath correlations are neglected. Thus, 
the overall error increases again. In between  there 
exists an ``optimal time step of least dependence", where the 
quantities are least sensitive to variations of $\dt$. This 
represents the ``principle of minimal sensitivity" for the optimal 
 choice of the time step $\Delta t$ for the QUAPI algorithm  
 (see also \cite{Golosov}). 
For $M$ finite and $K\rightarrow \infty$, the result would be 
independent of $\Delta t$ for small $\Delta t$ since the Trotter 
error would vanish and also the finite-memory error would not exist. 

The choice of $M$ and $K$ should be adapted to the chosen bath 
parameters. In the case of no driving, 
if the temperature is low, only few energy eigenstates 
are required, i.e.\ $M$ may be chosen small. However, low temperature 
induces long-range bath correlations. Therefore, the memory 
length $K$ has to be assumed large. The opposite holds true 
in the other limit of high 
temperature. In the case of driving, the number  $M$ of basis states 
is more important compared to the undriven case, since the variances 
oscillate strongly and higher lying energy states are excited. 
The memory length $K$ has to be 
reduced instead if one is interested in the oscillation amplitudes. 
However, for the mean value of the variances, the total memory 
length $K$ is again more important and should be maximized (see below).  

We shall choose two representative parameter sets for our 
considerations. Since the memory in Eq.\  (\ref{discr1}) 
is truncated in the QUAPI algorithm according to Eq.\  (\ref{discr2}), the 
crucial parameters are the temperature $T$ and the damping strength 
$\gamma$. The relatively high temperature $T=1.0$ and the small 
damping $\gamma=0.1$ form the first parameter set 
{\em (High temperature -- weak damping)}. For
 this choice, the numerical results are expected to agree well 
 with the analytic results because large $T$ suppresses the 
 long-time memory contributions in Eq.\  (\ref{discr1}) 
and additionally, a small $\gamma$ diminishes the influence of  the bath 
correlations (\ref{L}). 
 Our second parameter set  is given by  $T=0.1, \gamma=1.0$ 
 {\em (Low temperature -- strong damping)}. In this case, long-range 
 bath correlations (\ref{L}) play a major role and the truncation 
 of them will induce an error which will be larger than in the case of 
 a high temperature and weak damping. 
 For intermediate parameter 
 regimes, we find no qualitative differences. 
 
In all our calculations, we set $t_0=0$ and choose as 
the initial state the ground-state of the maximally curved 
(i.e\ $\omega_0^2 \rightarrow \omega_0^2 + 2 \epsilon)$ 
harmonic oscillator, i.e.\ $\rho_S(t_0 = 0) = |0\rangle \langle 0 |$. The 
corresponding initial variances in Eq.\  (\ref{final}) 
readily follow as $\sigma_{xx}^0 = 1 / (2 \sqrt{\omega_0^2 + 2 
\epsilon})$, $\sigma_{xp}^0 = 0$ and $\sigma_{pp}^0 = 
\sqrt{\omega_0^2 + 2\epsilon} / 2$. 
Our standard choice for the cut-off function will be the 
exponential cut-off 
(\ref{cut1}), if nothing else is stated. 
Furthermore, we always choose the dimensionless 
curvature $\omega_0=1.0$ in order to 
have a rather small separation of the energy levels in the undriven 
oscillator. This induces a high sensitivity 
on the number $M$ of basis states 
since the higher lying states are then easily populated thermally or 
by driving induced transitions. The choice of a larger $\omega_0$ would 
be more in favour of the numerical algorithm. 
%
%
\subsection{High temperature -- weak damping (no driving)}  
\label{hightempweakdamp}
\begin{figure}[th]
\begin{center}
\epsfig{figure=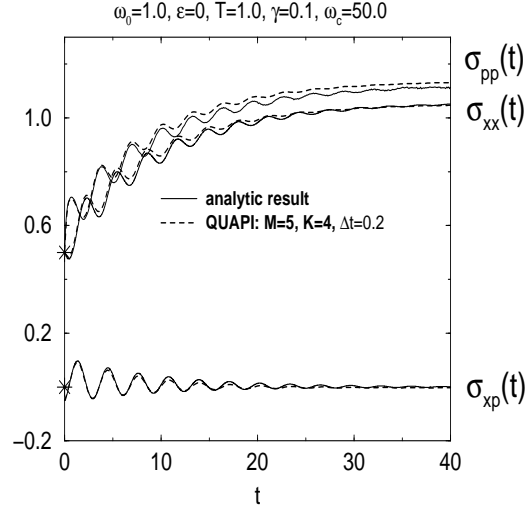,width=70mm,height=70mm,angle=0} 
\end{center}
\caption{Time-dependence of the variances $\sigma_{xx}(t), \sigma_{xp}(t)$,  
and $\sigma_{pp}(t)$ 
for the undriven dissipative quantum harmonic oscillator 
($\epsilon=0, \omega_0 =1.0$) with 
bath parameters $T=1.0, \gamma=0.1$ and an exponential cut-off Eq.\  
(\ref{cut1}) with $\omega_c=50.0$.  
In all the figures, we have used dimensionless quantities according 
to Eq.\  (\ref{scale}). The solid 
lines depict the analytic results (\ref{final}) while the dashed 
lines represent the numerical 
solution obtained by the  QUAPI  algorithm with $M=5, K=4, \dt=0.2
$. The asterisks mark the initial variances $\sigma^0_{xx}=\sigma^0_{pp} 
= 0.5$ and $\sigma^0_{xp}=0$.
 \label{fig1}}
\end{figure}

First, we consider the undriven case $\epsilon=0$. Fig.\ \ref{fig1} 
depicts the results for a high temperature $T=1.0$ and small 
friction $\gamma=0.1$. Here and in the following, we use the 
dimensionless quantities which have been introduced in  Eq.\  
 (\ref{scale}). Moreover, $\omega_0 =1.0$ and $\omega_c=50.0$. 
We find very good agreement with the analytic 
solution for the variances. The initial transient oscillations are 
reproduced and the asymptotic values for long times as well.  
The initial jump of $\sigma_{xp}(t)$ (of 
Eq.\  (\ref{inistep}))
is still visible, while the jump of $\sigma_{pp}(t)$ is proportional 
to $\gamma^2$ and is not visible on this plot. 

\begin{figure}[th]
\begin{center}
\epsfig{figure=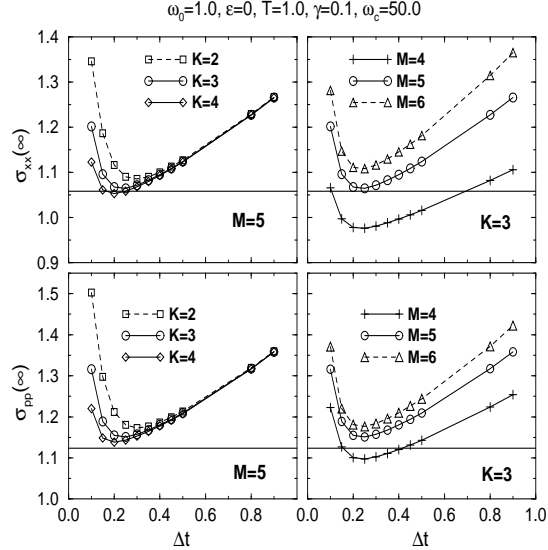,width=70mm,height=70mm,angle=0} 
\end{center}
\caption{Asymptotic values of the position (upper row) and 
momentum (lower row)  variances $\sigma_{xx}(\infty)$ 
and $\sigma_{pp}(\infty)$, respectively, as a function of the time-step $\dt$ and
 different combinations of the two QUAPI parameters $M$ 
(number of basis states) and $K$ (number of memory time steps) 
for the undriven dissipative quantum harmonic oscillator 
($\epsilon=0, \omega_0 =1.0$) and heat bath parameters $T=1.0, 
\gamma=0.1$ and $\omega_c=50.0$ with exponential cut-off Eq.\  (\ref{cut1}). 
For the left
 column figures, the number $M$ of basis states is fixed to $M=5$ and the 
 memory length $K$ is varied, 
 while for the right column figures, $K$ is fixed to $K=3$ and $M$ is varied.
 Interconnected symbols: solutions obtained by QUAPI.
 Horizontal solid line: analytic result (\ref{final}).
 \label{fig2}}
\end{figure}

To be able to study the dependence of the QUAPI algorithm on the
parameters $M, K$ and $\dt$ we consider the asymptotic 
values of the 
variances at long times. 
It is clear from Eq.\  (\ref{variances}) that $\sigma_{xp} 
(\infty)$=0, so we focus in Fig.\  \ref{fig2} 
on the two non-trivial variances 
$\sigma_{xx}(\infty)$ and $\sigma_{pp}(\infty)$. 
The qualitative dependence of both 
variances on the time step $\dt$ is always similar: The deviation 
increases with increasing $\dt$ due to the  
error proportional to $\dt^3$ in the Trotter splitting in Eq.\  
(\ref{split1}). For decreasing $\dt$, this error decreases  
and the ``finite-K"-error takes 
over. The relevant $\Delta t$-value on which we focus 
in the following is the one for which the numerical result 
varies the least, i.e.\ the minima in the curves in Fig.\  \ref{fig2} 
(``principle of minimal sensitivity" \cite{Golosov}). 

The left column of Fig.\  \ref{fig2} confirms that for a fixed memory 
length $\Delta t \cdot K$, a smaller time step $\Delta t$ induces a 
smaller Trotter-error whereas the finite-K-error remains roughly 
the same. 
While for a fixed $M$ (left column in Fig.\ \ref{fig2}) QUAPI tends 
to underestimate the analytic result as $K$ increases, a fixed $K$ 
and growing $M$ (right column) leads to an opposite trend, suggesting 
that indeed the analytic result will be approached best when {\it both} $M$ 
{\it and} $K$ become large (at a plateau $\Delta t$-value tending towards 
zero). 
\subsection{Low temperature -- strong damping}
\label{lowtempstrongdamp}
\subsubsection{No driving} \label{nodriving}
Fig.\  \ref{fig3} depicts the time-dependence of the variances 
in satisfactory agreement with the analytic result. 
The initial jumps of $\sigma_{xp}(t)$ and 
 of $\sigma_{pp}(t)$ are more pronounced in this 
 strong damping case since the jumps  are proportional 
to $\gamma$ and $\gamma^2$ (see Eq.\ (\ref{inistep})). 
The deviations in the transient behavior are due to the assumption 
of a strictly 
Ohmic classical dynamics (infinite cut-off $\omega_c$) in the 
analytic solution, see the discussion at the end of 
section  \ref{spddho}. 
They become more pronounced for low temperatures and strong 
friction because this assumption induces deviations in the short-time 
evolution of the variances on a time-scale $\omega_c^{-1}$. 
The bath-induced long-range memory at this low temperature   
carries the deviations over the whole range of the transient dynamics. 
The fact that the memory length $K$ is decisive for this low 
temperature is confirmed by the dashed-dotted line. 

The dependence of the asymptotic values 
$\sigma_{xx}(\infty)$ and 
$\sigma_{pp}(\infty)$ on the QUAPI parameters is shown in 
Fig.\  \ref{fig4}. The number $M$ of basis states is not so important,  
while the memory length $K$ is decisive. 
Again, the analytic prediction is correctly approached when both $M$ and 
$K$ are increased.
\subsubsection{With driving} \label{weakdriving}
Fig.\  \ref{fig5} demonstrates for a small driving amplitude reasonable 
agreement with the analytics. The long-memory parameter set 
with $K=6$ hits best the asymptotic mean value, but the oscillation 
amplitudes and frequencies are obtained best by the choice of a large $M=5$.  
In comparison to the undriven case 
($\epsilon =0$) the time averaged variances are almost 
unchanged (Figs.\ \ref{fig4} and \ref{fig6}) while the 
time-resolved behavior (Figs.\ \ref{fig3} and \ref{fig5}) 
displays notable differences. 

Fig.\ \ref{fig7} depicts the time evolution for the relatively 
large driving amplitude. As expected, for strong 
driving, a large number $M$ of basis states are required to 
describe the oscillations correctly. The averaged asymptotic values 
 $\overline{\sigma}_{xx}(\infty)$ and $\overline{\sigma}_{pp}(\infty)$ 
are plotted in Fig.\ \ref{fig8}.  Since the strong 
driving mixes  high energy eigenstates, 
the results are considerably more sensitive to the choice of $M$ than 
for weak driving (Fig.\  \ref{fig6} upper right panel). However, the 
same argumentation applies like in the undriven case (see Fig.\  \ref{fig4}). 
Considering the rather extreme parameters (small level-spacing, 
strong driving, low temperature, 
strong damping) the agreement with the analytic results is still satisfactory. 
\subsection{Diverging dynamics and dependence on the cut-off $\omega_c$} 
\label{diverging}
Fig.\ \ref{fig9} shows $\sigma_{xx}(t)$ for parameters 
belonging to an instability region of the 
Mathieu oscillator (\ref{mathieu}) \cite{mathieueq}, i.e.\  
the variances for 
the driven quantum harmonic oscillator diverge for long times. Since 
the QUAPI algorithm is restricted to a (finite) $M$-dimensional Hilbert 
subspace it cannot reproduce such an asymptotic divergence. 

\begin{figure}[th]
\begin{center}
\epsfig{figure=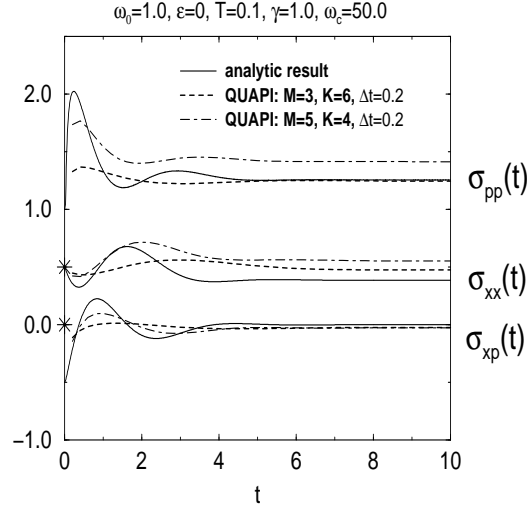,width=70mm,height=70mm,angle=0} 
\end{center}
\caption{Same as Fig.\  \ref{fig1}, but for the 
parameters $T=0.1, \gamma=1.0$. Here, the QUAPI parameters 
are $M=3, K=6, \dt=0.2$ (dashed line) and $M=5, K=4, \dt=0.2$ 
(dashed-dotted line). \label{fig3}}
\end{figure}

\begin{figure}[th]
\begin{center}
\epsfig{figure=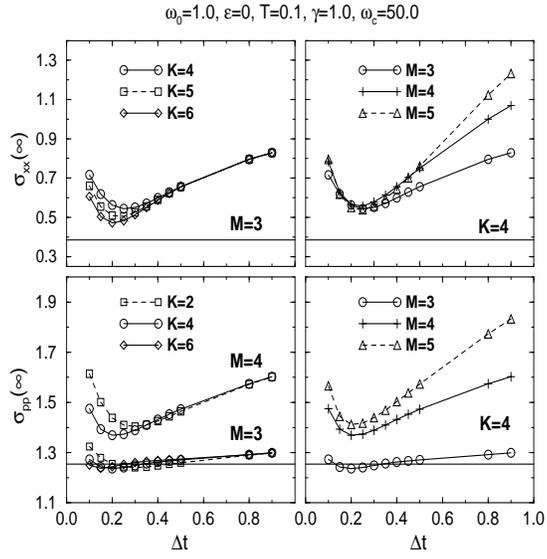,width=70mm,height=70mm,angle=0} 
\end{center}
\caption{Same as Fig.\  \ref{fig2}, but for the bath parameters 
$T=0.1, \gamma=1.0$. 
 \label{fig4}}
\end{figure}

\begin{figure}[th]
\begin{center}
\epsfig{figure=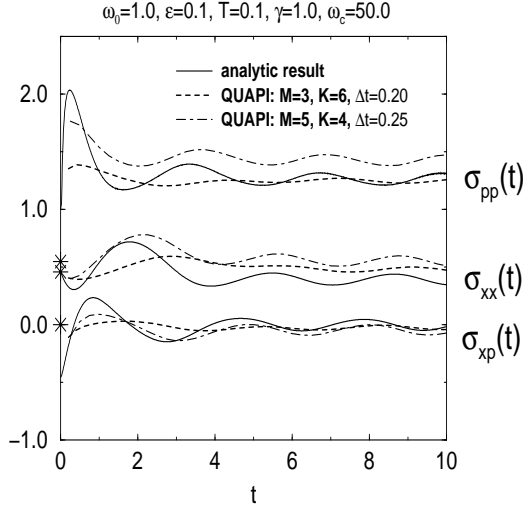,width=70mm,height=70mm,angle=0} 
\end{center}
\caption{Time-dependence of the variances $\sigma_{xx}(t), \sigma_{xp}(t)$ 
and $\sigma_{pp}(t)$ for the parametrically driven dissipative 
quantum harmonic oscillator with $\omega_0 =1.0$ and a small driving amplitude 
$\epsilon=0.1$. 
The bath parameters are $T=0.1, \gamma=1.0$ and $\omega_c=50.0$ 
(exponential cut-off Eq.\  (\ref{cut1})). 
The QUAPI parameters are $M=3, K=6, \dt=0.2$ (dashed line) 
and $M=5, K=4, \dt=0.25$ 
(dashed-dotted line). 
The asterisks mark the initial variances $\sigma^0_{xx}= 0.45, \sigma^0_{pp} 
=0.55$ and $\sigma^0_{xp}=0$.
\label{fig5}}
\end{figure}

\begin{figure}[th]
\begin{center}
\epsfig{figure=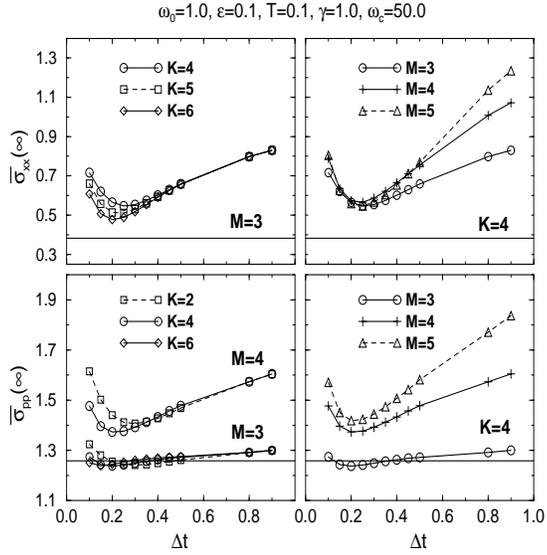,width=70mm,height=70mm,angle=0} 
\end{center}
\caption{Time-averaged asymptotic values of the position (upper row) and 
momentum (lower row)  variances $\overline{\sigma}_{xx}(\infty)$ 
and $\overline{\sigma}_{pp}(\infty)$, respectively, versus  
time-step $\dt$ for different combinations of the 
QUAPI parameters $M$ and $K$, small driving amplitude $\epsilon=0.1$ and 
bath parameters $T=0.1, \gamma=1.0, \omega_c=50.0$ 
(exponential cut-off Eq.\  (\ref{cut1})). 
 For the left column figures, the number $M$ of basis states is  
 fixed to $M=5$ and the memory length $K$ is varied, while for  
 the right column figures, $K$ is fixed to $K=3$ and $M$ is varied. 
 The oscillator frequency is $\omega_0=1.0$. 
 \label{fig6}}
\end{figure}

\begin{figure}[th]
\begin{center}
\epsfig{figure=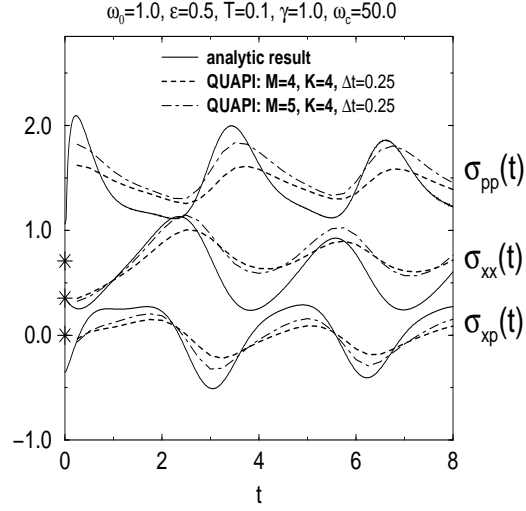,width=70mm,height=70mm,angle=0} 
\end{center}
\caption{Same as Fig.\  \ref{fig5}, but for the strongly driven  
case $\epsilon=0.5$. Here, the QUAPI parameters are 
$M=4, K=4, \dt=0.25$ (dashed line) and $M=5, K=4, \dt=0.25$ 
(dashed-dotted line). The asterisks mark the initial variances $\sigma^0_{xx}= 0.35, 
\sigma^0_{pp} 
=0.71$ and $\sigma^0_{xp}=0$.
\label{fig7}}
\end{figure}

\begin{figure}[th]
\begin{center}
\epsfig{figure=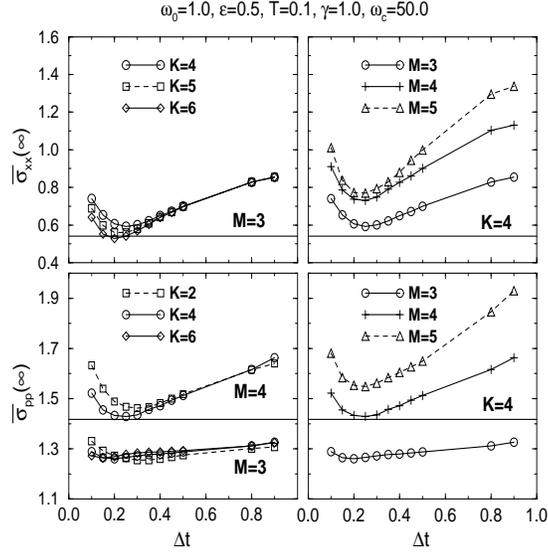,width=70mm,height=70mm,angle=0} 
\end{center}
\caption{Same as Fig.\  \ref{fig6}, but for the strongly driven 
case $\epsilon=0.5$.
 \label{fig8}}
\end{figure}

\begin{figure}[th]
\begin{center}
\epsfig{figure=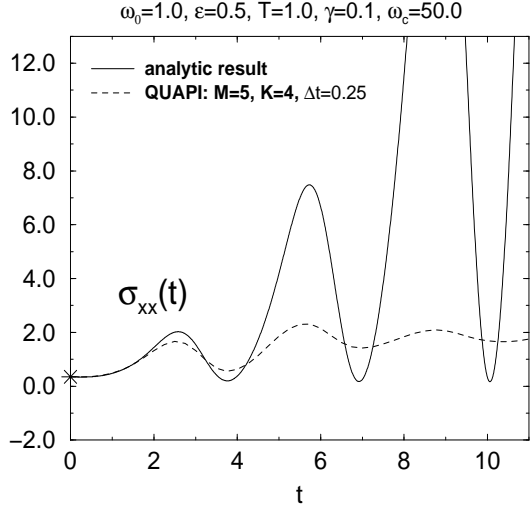,width=70mm,height=70mm,angle=0} 
\end{center}
\caption{Time-dependence of the position variance $\sigma_{xx}(t)$
 for a parameter set where the classical dynamics is 
 unstable, i.e.\ $\epsilon=0.5, \gamma=0.1$. The temperature 
 is $T=1.0$ and $\omega_c = 50.0$ (exponential cut-off Eq.\  (\ref{cut1})). 
The QUAPI parameters are $M=5, K=4, \dt=0.25$. 
The asterisk marks the initial variance $\sigma^0_{xx}= 0.35$. 
The oscillator frequency is $\omega_0=1.0$. \label{fig9}}
\end{figure}

\begin{figure}[th]
\begin{center}
\epsfig{figure=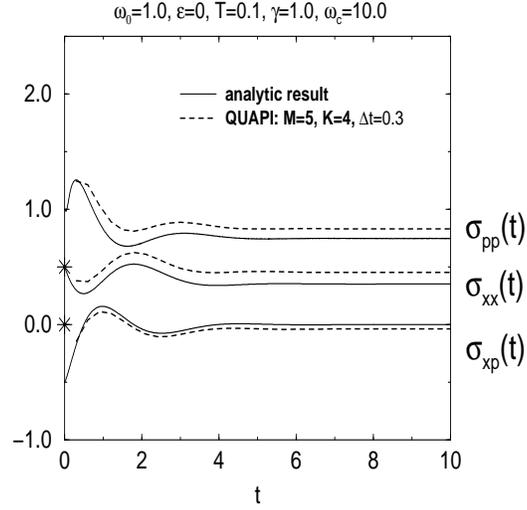,width=70mm,height=70mm,angle=0} 
\end{center}
\caption{Time-dependence of the variances $\sigma_{xx}(t), 
\sigma_{xp}(t)$ and $\sigma_{pp}(t)$
 for a small cut-off frequency $\omega_c = 10.0$ (exponential 
 cut-off Eq.\  (\ref{cut1})),  $\omega_0=1.0$, $\epsilon=0, \gamma=1.0$ and $T=0.1$. 
Here, the QUAPI parameters are $M=5, K=4, \dt=0.30$. 
The asterisks mark the initial variances $\sigma^0_{xx}=\sigma^0_{pp} 
= 0.5$ and $\sigma^0_{xp}=0$.\label{fig10}}
\end{figure}

The last issue we address is the dependence of the dynamics 
on the cut-off parameter $\omega_c$ and on the 
explicit {\it shape} of the cut-off function (\ref{cut1},\ref{cut2}). 
First, we 
keep an exponential cut-off but choose  
a smaller cut-off frequency $\omega_c$.  
 It is well known \cite{Weiss93} that for the (undriven) 
 quantum harmonic oscillator $\sigma_{pp}(\infty)$ diverges with 
  $\omega_c$,  while $\sigma_{xx}(\infty)$   
 is asymptotically 
 independent of $\omega_c$. In Fig.\  \ref{fig10}, we choose
 the ``worst'' case (i.e.\ low temperature and strong damping) 
  without driving 
 and decrease the cut-off to $\omega_c=10.0$. Compared to Fig.\ \ref{fig3}, 
 the value of $\sigma_{xx}(\infty)$ is indeed 
 practically unchanged while $\sigma_{pp}(\infty)$ has notably decreased. 
 
\begin{figure}[th]
\begin{center}
\epsfig{figure=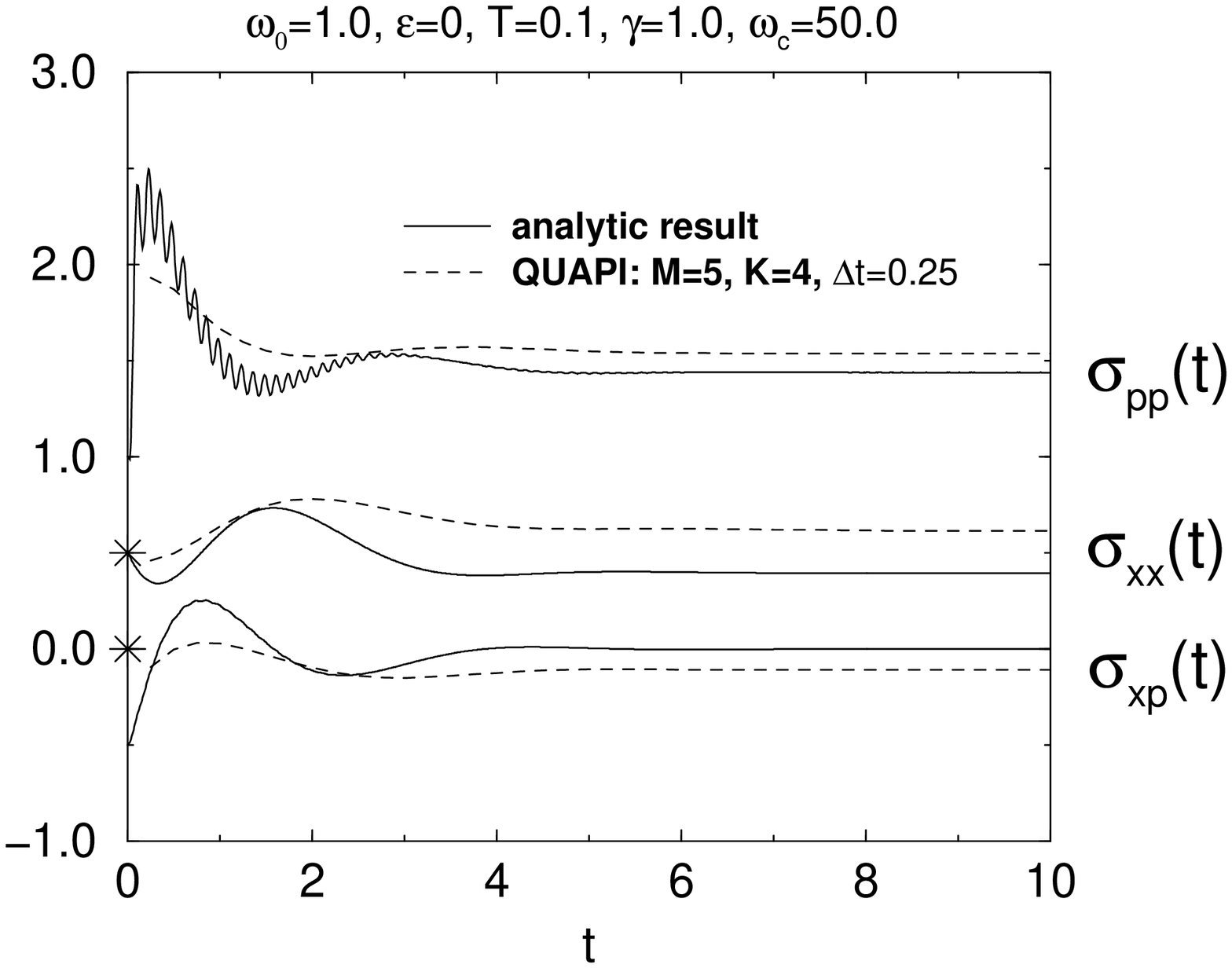,width=70mm,height=70mm,angle=0} 
\end{center}
\caption{Same as Fig.\ \ref{fig3}, but for a step-like cut-off Eq.\  
(\ref{cut2}) with $\omega_c=50.0$. Parameters are $\omega_0=1.0, 
\epsilon=0, T=0.1$ and $\gamma=1.0$. 
The QUAPI parameters are $M=5, K=4, \dt=0.25$. 
The asterisks mark the initial variances $\sigma^0_{xx}=\sigma^0_{pp} 
= 0.5$ and $\sigma^0_{xp}=0$.\label{fig11}}
\end{figure}

Fig.\ \ref{fig11} shows results for a step-like 
cut-off (\ref{cut2}). First, we observe that mainly the short-time 
behavior of 
the relaxation process is affected. Clearly, QUAPI with its 
restriction to only a few energy eigenstates 
cannot reproduce the transient high-frequency oscillations of 
$\sigma_{pp}(t)$. Second, we note that a step-like cut-off affects the 
decay of the response function $L(t)$ from Eq.\ (\ref{2.4})  
for $t \rightarrow \infty$. 
The real/imaginary part of $L(t)$ decays qualitatively like an 
algebraically damped cos/sin-function. While this might suggest 
a strong dependence of the QUAPI results on the 
 memory length,  we actually find a rather weak dependence since 
 the agreement between numeric and analytic results in Fig.\ \ref{fig11} is 
 not considerably worse than in Fig.\ \ref{fig3}. This 
 means that the memory truncation in Eq.\  (\ref{discr2}) is 
  in fact not very sensitive to the  
 choice  of the cut-off function $f_c(\omega,\omega_c)$ as long as one is not 
interested in the detailed short-time behavior.
\section{Conclusions} \label{conclusio}
We have studied the dependence of the QUAPI algorithm on its three 
numerical parameters, namely the  time-step $\dt$, the number $M$ 
of basis states,  and the memory length $K$. As a test system,  we 
have used the analytically solvable  dissipative quantum 
harmonic oscillator and its parametrically driven generalization. 
The comparison shows a decent agreement of the approximative 
numerical result with the analytic solution, even in the case 
with driving. This means that a 
{\it spatially continuous} system can be described reasonably well 
by taking only a few basis 
states and a finite memory length into account. 
For low temperatures and weak-to-moderate driving, the number $M$ 
of basis states has to  be chosen small and the memory length $K$ 
large, while in the opposite regime of high temperature, 
$M$ has to be large but $K$ may be chosen small. In both cases, 
satisfactorily large $M$ and $K$ values are still numerically feasible. 
For strong driving, the deviations increase but the QUAPI results are still 
in qualitative agreement with the analytic predictions. 

Our findings demonstrate the reliability of the QUAPI algorithm even in 
driven, {\it spatially continuous} systems and not only in finite, discrete 
dissipative quantum systems such as the  
spin-Boson-system. Therefore, the QUAPI algorithm may become a standard 
procedure for simulating open quantum systems in the presence of a 
novel class of time-dependent, not necessarily periodic driving 
fields. This technique is especially interesting for the study of  
decoherence in interacting two-level-systems processing  
quantum bits. There, the 
quantum gate operation prescribes the time-dependence 
of the external control fields which may exhibit a complex   
non-periodic time-dependence. 
\section*{Acknowledgement}
This work has been supported by the Deutsche Forschungsgemeinschaft Grant 
No.\ HA 1517/19-1 (P.H., M.T.), in part by the Sonderforschungsbereich 
 486 of the Deutsche Forschungsgemeinschaft (P.H.) and by the 
DFG-Graduiertenkolleg 283.
%

\begin{thebibliography}{10}
\bibitem{misprints}In eq.\ (60) in Ref.\ \cite{Zerbe95}, the factor 1/2 in the 
$\sigma_{xp}^0$-term has to be omitted. Moreover, in eq.\ (61), the last 
term should read correctly 
$...+2 \hbar m (\dot{f}_1^2 a_{11}+2\dot{f}_1a_{12} +a_{22})$.
\end{thebibliography}
\end{document}